\documentclass[aps,prl,twocolumn,superscriptaddress]{revtex4}
\usepackage{graphicx}
\usepackage{dcolumn}
\usepackage{bm}
\usepackage{verbatim}
\usepackage{color}
\usepackage[colorlinks]{hyperref}
\usepackage{amsmath}

\input{epsf}

\usepackage[sort&compress]{natbib}

\begin{document}
\title
  {Method for determining the residual electron- and hole- densities about the neutrality point over the gate-controlled  n $\leftrightarrow$ p transition in graphene}
\author{Ramesh G. Mani}
\affiliation{Department of Physics and Astronomy, Georgia State
University, Atlanta, GA 30303.}

\date{\today}

\begin{abstract}
 The Hall effect, and the diagonal resistance, which indicates a residual resistivity $\rho_{xx}
\approx h/4e^{2}$, are experimentally examined over the
p$\leftrightarrow$n transition about the nominal neutrality point in
chemical vapor deposition (CVD) grown graphene. A distribution of neutrality potentials is invoked in conjunction with multi-carrier conduction to model the experimental observations. From the modeling, we extract the effective residual electron- and hole- densities  around the nominal neutrality point. The results indicate mixed transport due to co-existing electrons and holes in large area zero-band gap CVD graphene devices, which indicates domain confined ambipolar currents broadly over the gate-induced n $\leftrightarrow$ p transition.


\end{abstract}
\maketitle
Graphene\cite{1,2} exhibits remarkable features such as massless
Dirac fermions, an anomalous Berry's phase, and signatures of
Hofstadter's spectrum,\cite{3,4,5,6} while allowing for the gate-induced carrier type conversion, from
holes to electrons, without crossing a
bandgap.\cite{7,8,9,10} When the Fermi level is placed at
the Dirac point in monolayer graphene, one expects the
\textit{density} of carriers, $n_{q}$, i.e., electrons ($n_e$) and
holes ($n_h$), to vanish, leading to a divergent Hall resistance,
$R_{xy}$, and a diverging resistivity $\rho_{xx}$, since $R_{xy}$
and $\rho_{xx}$ are  $\propto 1/n_{q}$. Remarkably, experiment has
reported a finite, nearly quantized $\rho = \sigma^{-1} \approx
h/4e^{2}$, reminiscent of the prediction $\sigma_{min} = e^2/h$
per massless Dirac channel,\cite{7,9} electron-hole
puddles,\cite{9,12b,14,13} along with Hall effect compensation.\cite{1} Here, we examine vanishing- instead of
diverging- Hall effect over the p$\leftrightarrow$n transition, in CVD
graphene,\cite{15} in addition to $R_{xy} \rightarrow 0$, as  the diagonal resistance through the Dirac peak indicates 
$\rho_{xx} \approx h/4e^{2}$ at the nominal neutrality point. We
reproduce these characteristics using a parabolic distribution $f(V_{N})$ of
neutrality potentials, $V_{N}$, in an ambipolar multi-conduction
model.\cite{16}  The results serve to extract the effective 
residual electron- and hole- densities at the nominal neutrality point, and as a function of the gate voltage around the neutrality point. The observed simultaneous non-vanishing electron- and hole- densities in this zero bandgap material at liquid helium temperatures suggest that the applied current could be simultaneously
carried by streams of electrons and holes within spatially separated current domains in these large area CVD graphene devices.

\begin{figure}[t]
\begin{center}
\includegraphics[width=60mm]{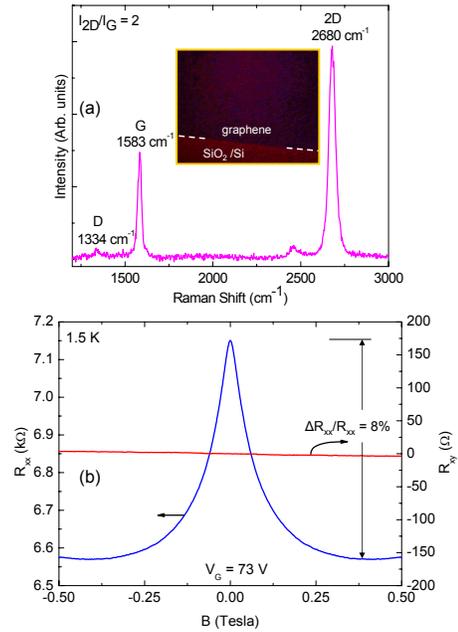}
\end{center}
\caption{(Color online) (a) The Raman intensity vs. the Raman shift for
graphene-on-$SiO_{2}/Si$-substrate. The D-, G-, and 2D- peaks have
been marked. The 2D-to-G intensity ratio, $I_{2D}/I_{G} = 2$,
suggests monolayer graphene. (b) The diagonal resistance,
$R_{xx}$, and the Hall resistance, $R_{xy}$, have been plotted vs.
the magnetic field, $B$, for a gate-voltage $V_{G} = 73$ V. The
figure indicates a relatively large ($ \approx 8$ per-cent) weak
localization correction. \label{fig: epsart}}
\end{figure}

\begin{figure}[h]
\begin{center}
\includegraphics[width=50mm]{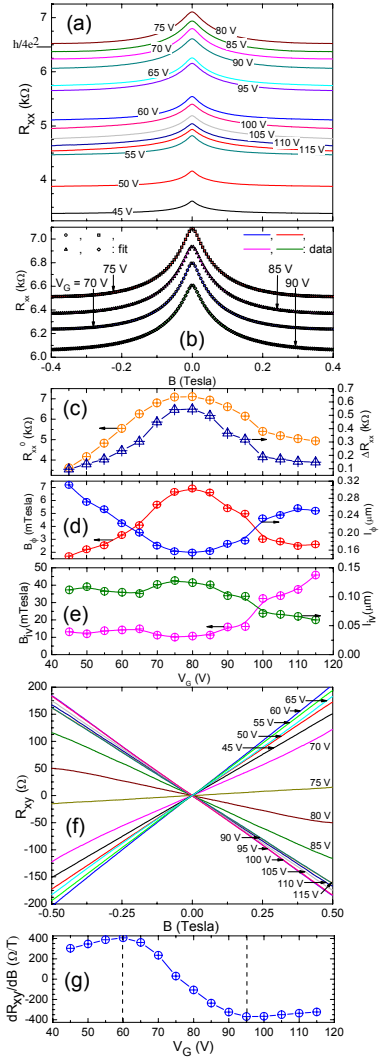}
\end{center}
\caption{ (Color online) (a) At $T=1.5$K, $R_{xx}$ is plotted vs. $B$ with $V_{G}$ as parameter. Here,  
$R_{xx}$ initially
increases up to $V_{G} = 75 $ V. For $V_{G} \ge 80$ V, $R_{xx}$
decreases with increasing $V_{G}$. (b) Fits to graphene weak
localization theory, see text.
(c) The fit extracted zero-field resistance, $R_{xx}^{0}$, and the
amplitude of the weak localization correction, $\Delta R_{xx}$,
are shown vs. $V_{G}$. (d) The fit extracted phase coherence
field, $B_{\phi}$ and the phase coherence length $l_{\phi}$ are
exhibited vs. $V_{G}$.  (e) The fit extracted inter-valley field,
$B_{iv}$ and the inter-valley length $l_{iv}$ are exhibited vs.
$V_{G}$. (f) At $T=1.5$K, the Hall resistance $R_{xy}$ is plotted
vs. $B$ with $V_{G}$ as a parameter. (g) Note the non-monotonic
variation in $dR_{xy}/dB$ vs. $V_{G}$. 
\label{afig: epsart}}
\end{figure}

\begin{figure}[t]
\begin{center}
\includegraphics[width=60mm]{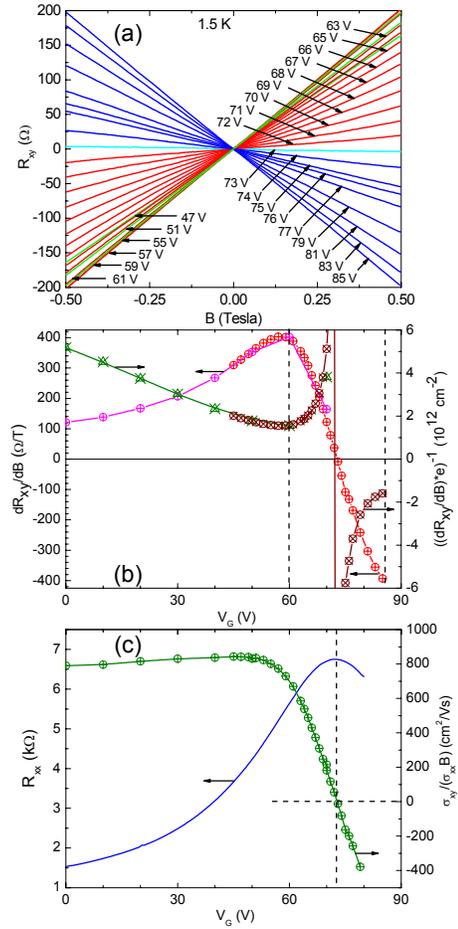}
\end{center}
\caption{ (Color online) (a) The Hall resistance, $R_{xy}$, is exhibited vs. $B$
for different $V_{G}$. The green (red) traces show a positive slope
$dR_{xy}/dB$ that increases (decreases) with $V_{G}$. The cyan
trace marks the boundary between the red and dark
blue traces. Dark blue traces show a negative slope. (b) $dR_{xy}/dB$ evaluated from
panel (a) are plotted as red symbols vs. $V_{G}$.  The green and
brown symbols show $1/((dR_{xy}/dB)e)$, the
carrier density in a single carrier model. (c) $R_{xx}$ (blue trace) peaks at the nominal neutrality
point (dashed vertical line). The evaluated
$\sigma_{xy}/(\sigma_{xx} B)$ (green symbols), would suggest a drop in the hole mobility over the range
$55 \le V_{G} \le 72$V. 
\label{figb:epsart}}
\end{figure}

\begin{figure}[t]
\begin{center}
\includegraphics[width=60mm]{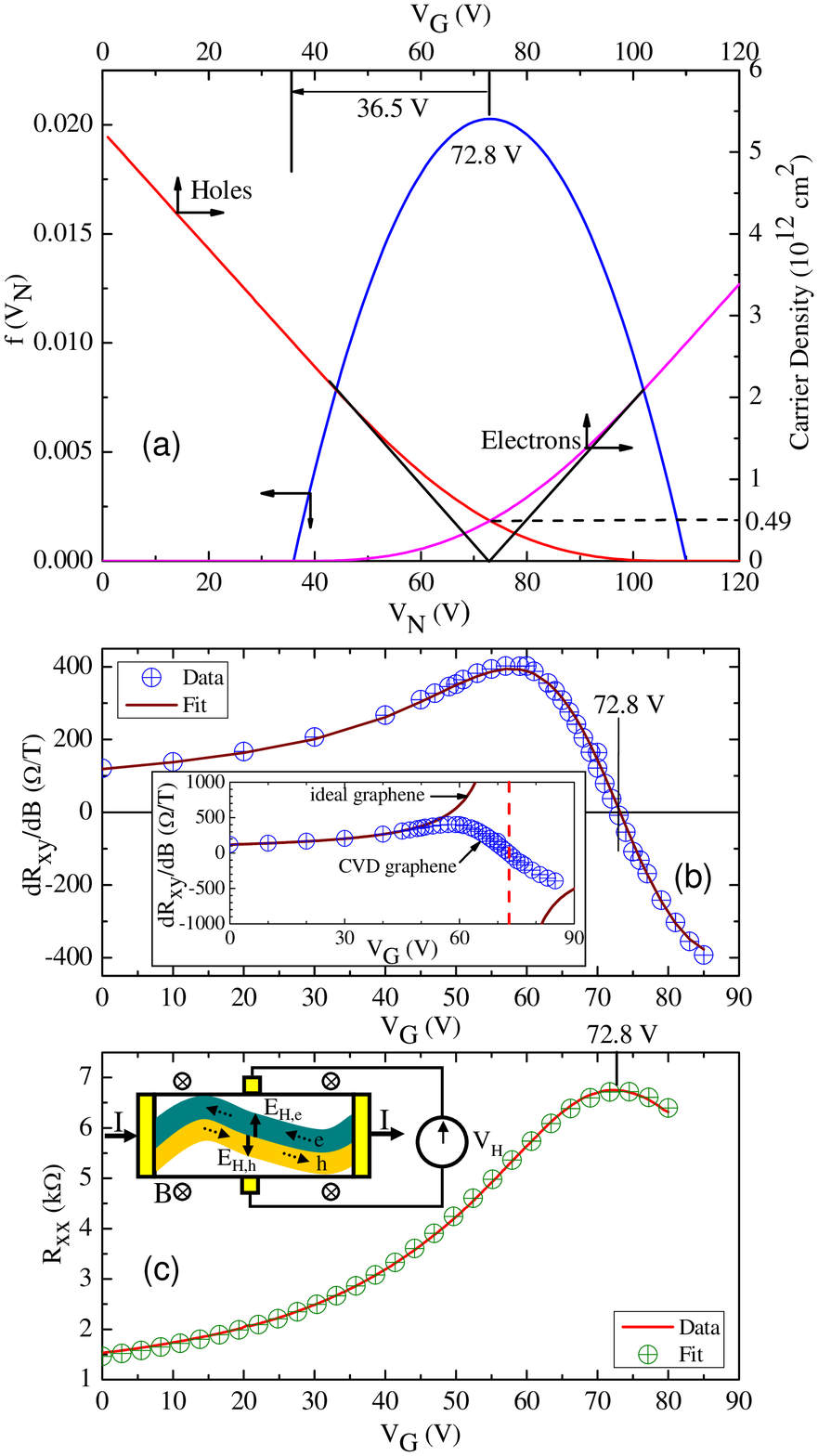}
\end{center}
\caption{ (Color online) (a) The blue trace shows the distribution function,
$f(V_{N})$, of the neutrality voltages. The red and magenta traces show the effective
hole and electron densities, respectively. The black
lines show the hole and electron densities in ideal
monolayer graphene. (b) The data (blue symbols) and best fit (brown
curve) of the $dR_{xy}/dB$ vs. $V_{G}$ obtained for the effective
carrier densities of panel (a). Inset: A comparison of
the $V_{G}$ dependence of $dR_{xy}/dB$ in ideal monolayer
graphene and CVD graphene. (c) The data (green symbols) and fit
(red curve) of the $R_{xx}$ vs. $V_{G}$ obtained for the effective
carrier densities of panel (a). Inset: Near the nominal
neutrality point, coexisting electron and hole streams, with
oppositely oriented Hall electric fields ($E_{H,e}$ and
$E_{H,h}$), carry the applied current. The Hall effect then
vanishes due to voltage compensation. \label{bfig: epsart}}
\end{figure}
Monolayer graphene was prepared using the CVD technique,\cite{15}
processed into mm-scale devices, and measured in a low temperature
magnet cryostat, see supplementary material.\cite{15b} The inset of Fig. 1(a) shows the graphene on $SiO_{2}/Si$, while
figure 1(a) exhibits the Raman scattering spectrum. Here, the
D-peak is small and the 2D-to-G intensity ratio, $I_{2D}/I_{G} =
2$, suggests monolayer graphene.\cite{15} Figure 1(b) shows the
diagonal resistance, $R_{xx}$, and $R_{xy}$ vs. the magnetic
field, $B$, at a back gate-voltage $V_{G} = 73$ V, close to the
nominal neutrality voltage, $\nu = 72.8 V$.
This panel (Fig. 1(b)) shows a large ($ \approx 8\%$) weak
localization correction to
$R_{xx}$.\cite{20,8,21,21b,22,9,10} The $|R_{xy}| = 3.7$
ohms at $B=0.5$ Tesla evaluates to $B/R_{xy}e = 8.5 \times
10^{13}$ cm$^{-2}$ at a $V_{G}$ where one expects $n_{e} < 1\times
10^{12}cm^{-2}$.

The next figures provide an experimental overview of the approach
to a vanishing Hall effect state at $V_G=\nu$. Figure 2(a) shows that $R_{xx}$ changes non-monotonically with
$V_{G}$, as $R_{xx}$ initially increases, followed by a decrease
above $V_{G} = 80 $ V. Here, $\nu \approx 77$ V. Also observable
in Fig. 2(a) is the weak localization correction which increases
towards $V_G = \nu$. Fig. 2(b) exhibits good data fits to weak
localization theory,\cite{21b} see supplementary \cite{15b}, on both the nominally hole- and nominally
electron-side  of $V_{G} = \nu$.  Fig.
2(c) shows the fit extracted zero-field resistance, $R_{xx}^{0}$,
and the weak localization amplitude, $\Delta R_{xx}$, vs. $V_{G}$.
At $V_{G}=75V$, $R_{xx}^{0} = 7.1k\Omega$ and $R_{xx} (0.4 T) =
6.5 k\Omega$ lie within $\approx 10 \%$ of $h/4e^{2} = 6.45
k\Omega$. Fig. 2(d) exhibits the fit extracted phase coherence
field $B_{\phi}$ and the phase coherence length $l_{\phi}$ vs.
$V_{G}$. The fit extracted inter-valley
field, $B_{iv}$ and the inter-valley length $l_{iv}$ are exhibited
vs. $V_{G}$ in Fig. 2(e). In sum, the fits work equally well on
either side of the nominal Dirac point, and the fit parameters,
exhibited in Fig. 2(c)-2(e), suggest increasing "localization"
with gate-induced carrier depletion.

Fig. 2(f) reports the Hall resistance. Here, the slope of $R_{xy}$ vs. $B$ is positive and
increases with $V_{G}$ up to $V_{G} = 60V$. From $60 \le V_{G} \le
75 V$, the slope remains positive although now it decreases in
magnitude with increasing $V_{G}$. Above $V_{G} = 75V$, the slope
turns negative and increases in magnitude up to $V_{G} = 100 V$.
Finally, the negative slope decreases in magnitude above $V_{G}
=100 V$. Fig. 2(g) shows the non-monotonic variation in
$dR_{xy}/dB$ vs. $V_{G}$, where $dR_{xy}/dB = \Delta R_{xy}/\Delta
B = R_{xy} (B)/B$ since $R_{xy}(0)
= 0$. Here, the dashed lines mark an anomalous band where the
$|dR_{xy}/dB|$ decreases in magnitude towards $V_G= \nu$ although
$n_q$ should decrease in this direction.

Characteristics over the crossover of the nominal neutrality point are
examined at smaller $V_{G}$ intervals in Fig. 3. In Fig. 3(a),
for the green traces, the positive slope $dR_{xy}/dB$ increases
with $V_{G}$. For the red traces, the positive slope decreases
with increasing $V_{G}$. The cyan trace, $V_{G} = 73 V$ marks a
boundary between the red and dark blue traces. [Data at this
$V_{G}$ appear in Fig. 1(b)]. For the dark blue traces, the slope
is negative and it increases in magnitude with increasing $V_{G}$.
$dR_{xy}/dB$ evaluated from Fig. 3(a) are plotted as red symbols
vs. $V_{G}$ in Fig. 3(b). Magenta symbols represent other data not
shown. Note that the $dR_{xy}/dB$ increases with increasing
$V_{G}$ only up to $V_{G} \approx 60V$. Above $V_{G} = 60 V$, the
slope decreases with increasing $V_{G}$ as it vanishes slightly
below $V_{G}= 73$ V. For a further increase in $V_{G}$,
$dR_{xy}/dB$ becomes negative and increases in magnitude with
$V_{G}$. Also shown  with the green and brown symbols in Fig. 3(b), is
$1/((dR_{xy}/dB)e)$, where $1/((dR_{xy}/dB)e)$ would equal the
carrier density for single carrier conduction. Here, $1/((dR_{xy}/dB)e)$ decreases up to $V_{G}
\approx 50$V as expected for hole depletion with increasing
positive $V_G$ via $n_{h} \approx C(\nu - V_{G})$ where $C \approx
7.2 \times 10^{-10} cm^{-2}/V$. The trace then deviates from
expectations as $1/((dR_{xy}/dB)e)$ increases with increasing
$V_{G}$ between $60 \le V_{G} \le 72$V, although one expects
further hole depletion. Similarly, from slightly below $V_{G} =
73$V, increase in $V_{G}$ should increase $n_e$, while the
$1/((dR_{xy}/dB)e)$ (brown symbols) indicates opposite behavior.

The $R_{xx}$ vs. $V_{G}$ data at $B=0$ are shown as the blue trace
in Fig. 3(c). The peak $R_{xx} \approx 6.5 k\Omega \approx
h/4e^{2}$ at $V_G = \nu$. We evaluate $ \sigma_{xy}/(\sigma_{xx}
B)$, where $\sigma_{xx}$ and $\sigma_{xy}$ are the conductivity
tensor components, and plot the result vs. $V_G$, see the green trace,
in Fig. 3(c). This plot suggests that holes in this specimen
exhibit an approximately constant mobility over the range $0 \le
V_{G} \le 50V$. Then, above $V_{G} = 50 V$,
$\sigma_{xy}/(\sigma_{xx} B)$ decreases with increasing $V_{G}$ as
it vanishes at $\nu$. At higher $V_{G}$, $\sigma_{xy}/(
\sigma_{xx} B)$ changes sign indicating electron conduction as it
increases in magnitude with $V_{G}$. Thus, a single carrier
interpretation which associates $\sigma_{xy}/( \sigma_{xx} B)$
with $\mu$ would suggest, unphysically, that the mobility changes with $V_G$.


Since a single carrier interpretation leads to un-physically large
$n_q$ close to the Dirac neutrality point, where $n_{q}$ ought
to vanish, and setting $\sigma_{xy}/(\sigma_{xx} B) = \mu$ also
leads to $\mu$ varying un-physically with $V_{G}$, see Fig. 3(c),
we considered, in the first iteration, ambipolar
conduction in a uniform-carrier-density, two carrier Drude model,
where at each $V_G$, there occurs both an electron and a hole
contribution at a constant density to the conductivities $\sigma_{xx}$ and $\sigma_{xy}$, see supplementary.\cite{15b}
Notably, this approach did not help to produce good fits of the data.

The large neutrality voltage, $V_N$, observed in CVD graphene on
$SiO_{2}/Si$ results from the combined gating effects of trapped
charge in the $SiO_2/Si$ substrate, charge donating impurities
below the graphene,\cite{31} and adsorbates on the graphene layer.
It appeared plausible that such disorder effects could combine to produce a
variation in the neutrality voltage across the mm-scale
CVD graphene specimen. Thus, we introduced a distribution
function, $f(V_{N})$, for $V_{N}$ with $\int f(V_{N})dV_{N} =1$, see supplementary.\cite{15b} Here, we focus upon the
results obtained with the parabolic distribution, $f(V_{N}) = A(w^{2} - (V_N - \nu)^2)$, blue curve
of Fig. 4(a), which is non-vanishing only between
$V_{N} = \nu \pm w$.

A distribution function $f(V_{N})$
introduces a gradation in the neutrality voltage, and a non-uniformity in the carrier density across the specimen. Suppose, for the sake of discussion, that the spread in neutrality voltages
occurs all in one direction, say, the x-direction, across the
width (W) of a Hall bar device of length L, with the left edge
of the Hall bar at $x=0$ is at $V_N = \nu -w$ and the right edge
of the Hall bar at $x=W$ is at $V_N = \nu + w$. Suppose the step
size in $V_{N}$ is $\Delta V_{N}$, then the spread in $V_N$ would
correspond to a number of strips of constant $V_N$, which
percolate from the bottom end of the Hall bar at $y=0$ to the top
end at $y=L$. (Scanning probe studies, see for example Fig. 3(a) of ref. \cite{13}, 
have suggested percolation of the hole and electron puddles over micron length scales). The spatial width, $\Delta x_{i}$, of these
percolating strips at a particular $V_{N}^{i}$ would be given by
$\Delta x_{i} =W f(V_N^{i}) \Delta V_{N}$. The total conductance
of the specimen system would equal the sum of
the conductances of all these strips. For simplicity, assume that
all the strips have the same mobility. Then, $\sigma_{xx} =
(e\mu/(1+(\mu B)^2)(1/W)(\sum n_{i} \Delta x_{i})$. In
neutrality voltage space, after separating out the
electron and hole components, one obtains $\sigma_{xx} =(e\mu/(1+(\mu
B)^2)[C\int_{\nu-w}^{V_G} (V_{G} - V_{N}) f(V_{N})dV_{N} +
C\int_{V_{G}}^{\nu + w} (V_{N} - V_{G}) f(V_{N})dV_{N}]$.
Similarly, $\sigma_{xy} =(e\mu^{2} B/(1+(\mu
B)^2)[-C\int_{\nu-w}^{V_G} (V_{G} - V_{N}) f(V_{N})dV_{N} +
C\int_{V_{G}}^{\nu + w} (V_{N} - V_{G}) f(V_{N})dV_{N}]$. 

Using this model, we evaluated the diagonal resistance $R_{xx} =
\rho_{xx} (L/W)$ and $dR_{xy}/dB = (R_{xy} (0.5 T)/(0.5 T))$. A
least squares data-fit was carried out with $\nu$ and $w$ as the
fitting parameters. The blue trace in Fig. 4(a) illustrates the $f(V_N)$ that served to
obtain the best data fit shown in Fig. 4(b) and Fig. 4(c). This
$f(V_{N})$ peaks at $\nu \approx 72.8V$ and exhibits a width
$w=36.5 V$. The extracted effective electron and hole densities for the best
fit $f(V_N)$ are also shown in Fig. 4(a). Here, the effective hole density
at a particular gate voltage $V_{G} \le \nu+w$ is $n_h^{eff.} (V_{G}) =
C\int_{V_{G}}^{\nu+w} (V_{N}-V_{G})f(V_N) dV_{N}$. Similarly, the effective electron density at  $V_{G} \ge \nu-w$ is $n_e^{eff.}(V_{G}) = C\int_{\nu -
w}^{V_G} (V_{G}-V_{N})f(V_N) dV_{N}$. The notable features are that (a) for a broad range of
$V_G$ about $\nu = 72.8 V$, one finds non-vanishing $n_e^{eff.}$ and
$n_h^{eff.}$, unlike the expectations for $n_{e}$ and $n_{h}$ indicated by the black lines for
ideal graphene, and (b) at $\nu = 72.8 V$, the effective densities are given by $n_h^{eff.} = n_e^{eff.} = 0.49
\times 10^{12} cm^{-2}$. Thus, this approach helps us to determine the effective residual densities of electrons and holes at the nominal neutrality point. Fig. 4(b) shows the experimental $dR_{xy}/dB$ data along with the
best fit, while Fig. 4(c) shows $R_{xx}$ data along with the best
fit. These good fits suggest that the model captures the
transport behavior observed in the measurements.


When the Fermi level in ideal monolayer graphene is brought
towards the Dirac neutrality point, naively, one expects $n_q$ to
progressively vanish, leading to a divergent $dR_{xy}/dB$, see
brown trace Fig. 4(b) (inset), with a discontinuity (dashed red
line in Fig. 4(b), inset) in $dR_{xy}/dB$ vs. $V_{G}$ at the
nominal neutrality point.
Measurements for large area disordered CVD graphene devices shown here illustrate  that the
$dR_{xy}/dB$ changes smoothly from positive to negative values
over a broad range of $V_G$, and $dR_{xy}/dB = 0$ at $V_G=\nu$.
Indeed, in experiment, one might set $V_{G} = \nu$ and obtain
vanishing Hall effect, or close to it, as in Fig. 1, with stable and reproducible results.

This modelling suggests ambipolar transport over a wide range of
$V_{G}$, about $V_G = \nu = 72.8V$, see Fig. 4(a), because there is a concurrent non-vanishing effective electron- and hole- density over a broad span of $V_{G}$. The model
provided here is reminiscent of the uniform-carrier-density two
carrier Drude model. However, this model differs
on the point that instead of two carriers at fixed densities,
there is a gradation in the density of both the electrons and the
holes over the sample. If we set the effective hole density to the hole density, i.e., $n_{h}^{eff.} = n_{h}$, and the effective electron density to the electron density, i.e., $n_{e}^{eff.} = n_{e}$, then this model is formally similar to the two carrier
Drude model. Thus, we can use the
predictions of the two carrier ambipolar Drude model to understand
some observations. For example, in the ambipolar two carrier model
with $\mu_h = \mu_e = \mu$ and $\mu B <<1$, $dR_{xy}/dB = (1/e)
(n_h-n_e)/(n_h+n_e)^2$. Thus, $dR_{xy}/dB \propto n_h-n_e$, and
$dR_{xy}/dB$ should vanish at $V_G=\nu$, where $n_h = n_e$, as observed in experiment, see Fig. 3(b) or
Fig. 4(b). A qualitative picture, which suggests streams rather than puddles, for the $V_{G} = \nu$ condition
is exhibited in Fig. 4(c), inset. At $V_G=\nu$, the applied
current is equally split by symmetry between
counter-propagating holes and electrons. In a magnetic field, the
Hall electric field, $E_H$, within these hole- and electron-
current domains will be oppositely directed, which will lead to a
vanishing global Hall effect.\cite{24,23} Yet, the diagonal
resistance/resistivity would remain finite due to the finite residual densities and the non-vanishing mobility.  For the experimental parameters, this residual
resistivity turns out to be $\approx h/4e^{2}$.\cite{9} Here, the
observed weak temperature dependence of the residual resistivity
is attributed to weak localization and the relative temperature
insensitivity of the transport parameters at the nominal
neutrality point.\cite{30} Fig. 4(c), inset, also helps to qualitatively understand small deviations from
$V_G=\nu$:  Deviation towards the hole regime, i.e, $V_G \le \nu$,
will increase the hole current at the expense of the electron
current. Then, the hole Hall electric field will exceed the
electron Hall electric field, yielding a net hole-like Hall
effect. On the other hand, deviating towards the electron regime,
i.e., $V_{G} \ge \nu$, will increase the electron current at the
expense of the hole current, yielding a net electron-like Hall
effect.

\pagebreak

\end{document}